\documentclass[twocolumn,bibnotes,showpacs]{revtex4-1}

\usepackage{graphicx}
\usepackage{color}
\usepackage{soul} 
\usepackage{amsmath}
\usepackage{hyperref}
\usepackage{braket}
\usepackage{mathtools}
\usepackage{lipsum}
\hypersetup{colorlinks=true, citecolor=cyan, urlcolor=blue, linkcolor=blue}
\usepackage[capitalize]{cleveref}

\newcommand{\RMP}[4]{\textit{#1}, Rev. Mod. Phys. \textbf{#2}, #3 (#4)}

\newcommand{\PRL}[4]{\textit{#1}, Phys. Rev. Lett. \textbf{#2}, #3 (#4)}
\newcommand{\PRR}[4]{\textit{#1}, Phys. Rev. Research. \textbf{#2}, #3 (#4)}
\newcommand{\PRA}[4]{\textit{#1}, Phys. Rev. A \textbf{#2}, #3 (#4)}
\newcommand{\PRB}[4]{\textit{#1}, Phys. Rev. B \textbf{#2}, #3 (#4)}
\newcommand{\PRApplied}[4]{\textit{#1}, Phys. Rev. Appl. \textbf{#2}, #3 (#4)}
\newcommand{\PRX}[4]{\textit{#1}, Phys. Rev. X \textbf{#2}, #3 (#4)}
\newcommand{\Science}[4]{\textit{#1}, Science \textbf{#2}, #3 (#4)}
\newcommand{\Nature}[4]{\textit{#1}, Nature \textbf{#2}, #3 (#4)}
\newcommand{\NPhys}[4]{\textit{#1}, Nat. Phys \textbf{#2}, #3 (#4)}

\newcommand{\NJP}[4]{\textit{#1}, New J. Phys. \textbf{#2}, #3 (#4)}
\newcommand{\SciRep}[4]{\textit{#1}, Sci. Rep. \textbf{#2}, #3 (#4)}

\begin{document}

\title{Two-photon sideband interaction in a driven quantum Rabi model : Quantitative discussions with derived longitudinal drives and beyond the rotating wave approximation}

\author{Byoung-moo Ann}
\email{byoungmoo.ann@gmail.com}
\thanks{Current address : Quantum Technology Institute, Korea Research Institute of Standards and Science, 34113 Daejeon, South Korea}
\author{Wouter Kessels}
\author{Gary A. Steele}
\affiliation{Kavli Institute of Nanoscience, Delft University of Technology, 
2628 CJ Delft, The Netherlands} 
\date{\today}

\begin{abstract}
In this work, we analytically and numerically study the sideband interaction dynamics of the driven quantum Rabi model (QRM).
We focus in particular on the conditions when the external transverse drive fields induce first-order sideband interactions.
Inducing sideband interactions between two different systems is an essential technique for various physical models, including the QRM.
However, despite its importance, a precise analytical study has not been reported yet that successfully explains the sideband interaction rates in a driven QRM applicable for all system parameter configurations. 
In our study, we analytically derive the sideband interaction rates based on second-order perturbation theory, not relying on the rotating wave approximation (RWA). Our formula are valid for all ranges of drive frequencies and system's parameters.
Our analytical derived formula agrees well with the numerical results in a regime of moderate drive amplitudes. 
Interestingly, we have found a non-trivial longitudinal drive effect derived from the transverse drive Hamiltonian. This accounts for significant corrections to the sideband interaction rates that are expected without considering the derived longitudinal effect.
Using this approach, one can precisely estimate the sideband interaction rates in the QRM not confining themselves within specific parameter regimes for moderate drive amplitudes.
This provides important contributions for quantitatively understanding experiments described by the driven QRM.
\end{abstract}

\maketitle
\section{Introduction}
The quantum Rabi model (QRM)\cite{QRM} constitutes the essence of the light-matter interactions at the quantum level.
It specifically describes the interaction between a two-level system (qubit) and a single cavity mode.  
The QRM has been extensively studied both for fundamental interest and for applications in quantum information processing.
In addition, the QRM can describe many systems. It was originally formulated to mathematically describe cavity quantum electrodynamics (QED), and study the interaction between a trapped atom and cavity mode. Beyond atomic physics, it can also be extended to any other systems that have an analogy with the cavity-QED, such as quantum-dots in microcavities and various types of qubits that are transversely coupled to superconducting cavities. 
Moreover, the extended versions of the QRM have been widely investigated \cite{QRM1,QRM2,QRM3}. 

The question of how to implement in-situ tuneable state transfer between the qubit and cavity mode (sideband interactions) is an important aspect of studying the QRM. In particular, it is crucial for quantum gate operation using qubits and can be employed for quantum state engineering of the cavity. There are several ways to achieve this. 
One approach is to suddenly switch the transition frequency of the qubit ($\omega_q$).
If the qubit is initially far-off resonant from the cavity transition frequency ($|\omega_q-\omega_c|\gg 1$),
then we consider the qubit to be isolated and they are effectively uncoupled. 
However, if the qubit's transition frequency jumps from $\omega_q$ to $\omega_c$, then the qubit and cavity become resonant and  coherent state transfer begins.
Consequently, by shifting the $\omega_q$, we can turn the interaction between the qubit and the cavity on and off.
The other approach is to parametrically modulate the qubit's transition frequency. 
The first order sideband interactions between the qubit and cavity occur when the modulation frequency $\omega_m$ satisfies the matching conditions($\omega_m=|\omega_q\pm k\omega_c|$, $k$ is integer).

These approaches require that the frequency of the qubit should be tunable over short time scales.
This is technically feasible if one employs superconducting qubits with SQUID loops and on-chip magnetic flux lines.
The sudden frequency switch was realized in \cite{Fock}, where the authors create Fock states in a superconducting cavity.
Inducing the first-order sideband interactions by flux modulation was proposed in \cite{Beaudoin-PRA-2012} and was experimentally implemented in \cite{Strand-PRB-2013}.
In all cases, the systems can be modeled by the QRM.
Although these cases successfully demonstrate the state transfer from the qubit to the cavity, introducing tuneability into the qubit's transition frequency leads to another side-effect: pure dephasing induced by external noise. 
For example, when the tuneability relies on the magnetic flux through the squid loops, then the magnetic field noise into the loops accounts for the qubit's pure dephasing.

One can also induce the sideband interactions without any frequency tuneability of the qubit and cavity by applying the external transverse drive at the proper frequencies. 
This scheme is implementable with a fixed frequency qubit. Therefore, the system is insensitive to the external noise and the qubit's dephasing rate is only limited by the qubit's decay rate \cite{ann-PRA-2020}.
For the first-order sideband interaction in the QRM, which is typically the most attractive type,
the transition is unfortunately dipole forbidden and therefore only a two (or any higher even number) photon drive can induce the transition. The description of the selection rule of the QRM is well explained in the Appendix E of Ref.~\cite{Blais-PPA-2006}.
This complicates the analytical solution for the interaction rates because we cannot capture the transitions simply by first-order perturbation theory.

In this paper, we perform a quantitative study of the first-order sideband interaction in the QRM induced by two-photon transverse drive fields.
We analytically derive the interaction rates based on perturbative calculation up-to second order without relying on the rotating wave approximations \cite{RWA} in the Hamiltonian. We specifically investigate the parameter regimes that are familiar in circuit quantum electrodynamics (QED) experiments.
In circuit QED, the frequency matching condition for sideband interactions often requires drive parameters that are beyond the rotating wave approximation (RWA) \cite{Ann-PRR-2021}, 
and therefore one should not rely on the RWA in the analytical derivation of the interaction rates.
Moreover, under the transverse drive field, the qubit's frequency should be modulating at the lab frame, which effectively amounts to the longitudinal drive. This effect was typically neglected although it also can induce the sideband interactions. 

Whereas a number of studies have examined how the external transverse drive fields affect the qubits or similar systems beyond the RWA \cite{Andrews-JPB-1975,Werlang-PPA-2008,Fuchs-Science-2009,Tuorila-PRL-2015,BSS-PRB-2017}, and 
there are a few studies quantitatively discussing the sideband interaction rates between the qubits and cavities \cite{Blais-PPA-2006,Deppe-NPhys-2008, Zeytinoglu-PPA-2016,Chen-PPA-2016,Krinner-PRApplied-2020,Ann-PRR-2021},
a satisfactory quantitative study of the driven QRM beyond the RWA regime and considering the derived longitudinal drive effect has not yet been reported.
Although the quantitative work on drive-induced sideband interaction rates beyond the RWA is presented in several papers including our previous work \cite{Krinner-PRApplied-2020, Ann-PRR-2021}, these are relevant with transmon \cite{Koch-PRA-2007} coupled to a resonator, not described by the QRM due to their weakly nonlinear nature. Considering recently rising interests in strongly anharmonic systems such as Fluxonium qubits \cite{Nguyen-prx-2019}, and spin qubits \cite{Samkharadze-science-2018} whose interface to a cavity can be modeled by the QRM, extending the discussion beyond transmons systems should deserve large attention.

To our best knowledge, the initial attempt to analytically derive the two-photon sideband interaction rates in the QRM was given in \cite{Blais-PPA-2006}. In that study, a charge qubit device dispersively coupled to the cavity was modeled by the QRM. 
However, the analytically derived interaction rates are significantly smaller than the simulation results. 
In our work, we found that the RWA significantly distorts the calculated sideband interaction rates for some system parameters. 
We also investigate if the transverse drive field accounts for a derived longitudinal drive effect, which also significantly contributes to the total sideband interaction rates.
Our analytical predictions of the frequency matching conditions and sideband interaction rates are well consistent with the numerical  results when we have moderate drive amplitudes. 
Although our analytical model fails to explain the sideband interaction rates as the drive strength becomes comparable to the detuning between the qubit and drive, it nonetheless yields more precise predictions in general than the previous analytical model. It is crucial to remark that we find quantitative and qualitative differences between the QRM and transmon model cases both in the qubit frequency shifts and sideband interaction rates, which will be also discussed in the main part of this paper.

This paper is organized as follows. In Sec.~\ref{theory}, we analytically derive the expected matching frequencies and sideband interaction rates based on the perturbation theory up to second-order. A description of the numerical simulation performed in this study is given in Sec.~\ref{simulation}. We compare the analytical and numerical calculation results in Sec.~\ref{Results} with extensive parameter scanning. We also discuss the validity and limitation of our theory in this section.   
Finally, we conclude our paper in Sec.~\ref{conclusion}.

\section{Theoretical description}
\begin{figure}
    \centering
    \includegraphics[width=1\columnwidth]{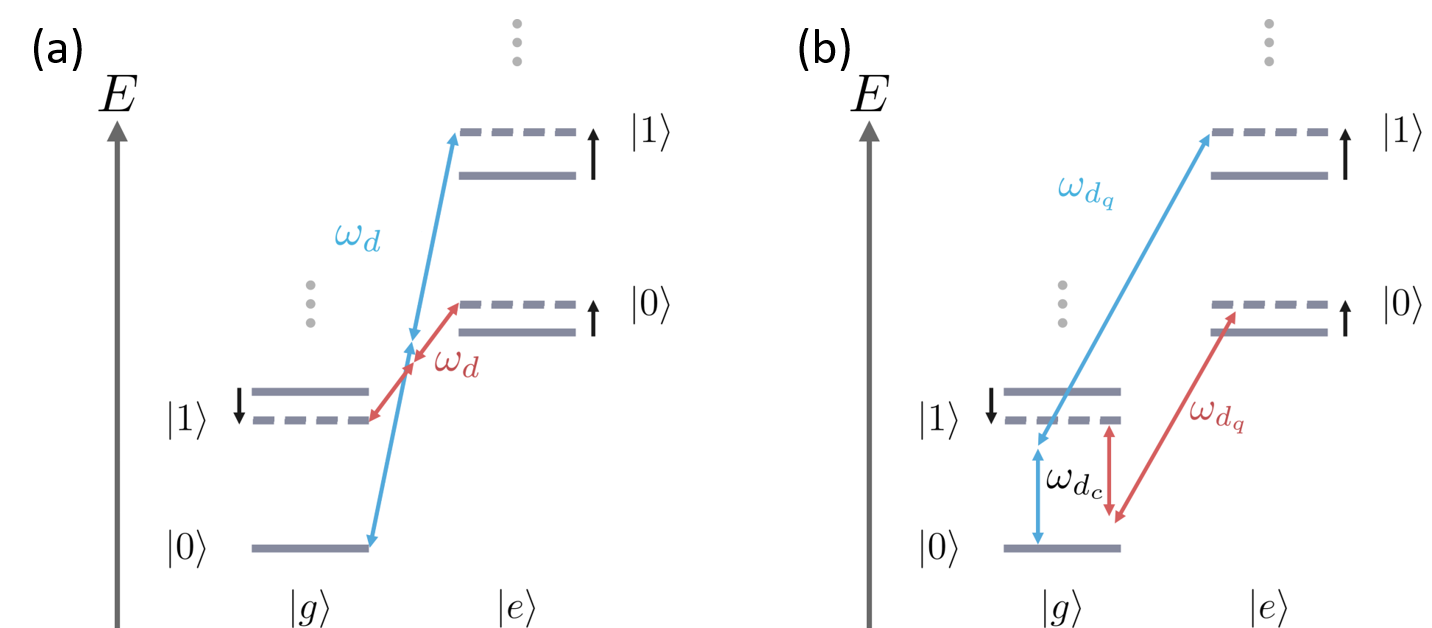}
    \caption{Descriptions of the first order red and blue sideband interactions in quantum Rabi model (QRM). The arrows indicate the external drives that satisfy the matching conditions for red and blue sideband interactions (red and blue arrow, respectively). A two-photon drive is required because the first order sideband interactions in QRM are dipole forbidden. (a) Single frequency (monochromatic) drive. (b) Two-frequency (bi-chromatic) drive.
    $\ket{gn}$ and $\ket{en}$ (corresponding to dashed lines) represent the dressed states of the system. The black arrows indicate the frequency shifts induced by the external drive fields and the qubit-cavity bare coupling $g$. The bare states are depicted by solid lines.}
    \label{fig:energylevels}
\end{figure}

In this section, we derive an analytical formula to predict the matching frequencies and sideband interaction rates. We investigate two possible schemes for the first order two-photon sideband interactions in the QRM, as shown in Fig.~\ref{fig:energylevels}. The possible scheme for the first order sideband interactions is described in Fig.~\ref{fig:energylevels}(a), where the drive field has only a single frequency component $\omega_d$ (monochromatic drive). The downside of this approach is that there is no flexibility in choosing the drive frequency for the given qubit and resonator frequencies. When using two different drive frequencies (i.e., a bi-chromatic drive), we can have more flexibility in choosing the drive frequencies. Fig.~\ref{fig:energylevels}(b) describes the bi-chromatic drive case. One drive frequency is close to the resonator ($\omega_{dc}$, which is called a resonator friendly drive in this paper), and the other  is close to the qubit ($\omega_{dq}$, qubit friendly). The solid line and dashed lines refer to the bare and dressed energy states of the QRM, respectively.

\label{theory}
\subsection{Schrieffer-Wolff transformation}
The transversely driven QRM Hamiltonian reads,
\begin{equation}
\label{eq:qrm}
\begin{split}
    \hat{H} =
    \underbrace{\frac{{\omega}_q}{2} \hat{\sigma}_z + {\omega_c} \hat{a}^\dagger \hat{a} + g(\hat{a}^\dagger+\hat{a})\hat{\sigma}_x}_{\hat{H}_\textup{QRM}} \\ + \underbrace{\sum_{i}\Omega_{d}^{(i)}\hat{\sigma}_x\cos{(\omega_{d}^{(i)}t)}}_{\hat{H}_\textup{drive}}&.
\end{split}
\end{equation}
Here, $\hat{\sigma}_{z,x}$ is the z and x components of the Pauli operators, and $\hat{a}$ is the cavity field operator. $\omega_{q,c}$ are angular frequencies of the qubit and cavity, respectively. $\Omega_{d}^{(i)}$ and $\omega_{d}^{(i)}$ refer to $i$-th component of the drive amplitude and frequency. It is also useful to define $\epsilon_d^{(i)} = \Omega_d^{(i)}/2$ the drive strength for use later in this paper. We are interested in the dispersive coupling regime where $|\omega_{q}-\omega_{c}| \gg g$. We are also interested in those drive frequencies $\omega_{d}^{(i)}$ that are far off resonant to $\omega_{q,c}$, and those drive amplitudes $\Omega_{d}^{(i)}$ that are smaller than $|\omega_{q,c}-\omega_{d}|$.
With these parameter conditions, $\hat{H}_d$ can be considered as a perturbation to $\hat{H}_\textup{QRM}$. We can then perturbatively diagonalize the $\hat{H}=\hat{H}_\textup{QRM}+\hat{H}_\textup{drive}$ using Schriffer-Wollf transformation \cite{SW}. The transform operator $\hat{U}$ takes a form of $\hat{U} = \textup{exp}(\beta^{*}\hat{\sigma}_{+}-\beta\hat{\sigma}_{-})$. We define $\hat{X} = \beta\hat{\sigma}_{-}-\beta^{*}\hat{\sigma}_{+}$ in the following.  
The transformed Hamiltonian $\hat{H'}$ is given by,
\begin{equation}
\label{eq:transform}
\begin{split}
    \hat{{H'}} = {\hat{U}\hat{H}\hat{U}^\dagger}+i(\partial_{t}\hat{U})\hat{U}^\dagger.
\end{split}
\end{equation}
The first term in Eq.~\ref{eq:transform} can be calculated using the Hausdorff expansion \cite{Hausdorff},
\begin{equation}
\label{eq:Hausdorff}
\begin{split}
    e^{\lambda\hat{X}}\hat{{H}}e^{-\lambda\hat{X}} =
    \hat{H} - \lambda[\hat{H},\hat{X}] + \frac{\lambda^2}{2}[[\hat{H},\hat{X}],\hat{X}] + \cdot \cdot \cdot
\end{split}
\end{equation}
When $\beta \ll 1$, we can truncate the expansion to the low order of $\lambda$. 
To capture the two-photon transitions, we should include at least to the second order of $\lambda$. 
Meanwhile, the second term in Eq.~\ref{eq:transform} can be approximated by \cite{Blais-PPA-2006}, 
\begin{equation}
\label{eq:time}
\begin{split}
    (i\partial_{t}\hat{U})\hat{U}^\dagger \approx \\ 
    \frac{i}{2}&(\beta^{*}\dot{\beta}-\beta\dot{\beta}^{*})\hat{\sigma}_{z} + i(\dot{\beta}\hat{\sigma}_{-}-\dot{\beta}^{*}\hat{\sigma}_{+}).
\end{split}
\end{equation}
$\hat{H'}$ is then expressed by,
\begin{widetext}
\begin{equation}
\label{eq5}
\begin{split}
    \hat{H'}&\approx \underbrace{\frac{\omega_{q}}{2}\hat{\sigma_z} + \omega_{c}\hat{a}^\dagger\hat{a} + g(\hat{a}+\hat{a}^\dagger)\hat{\sigma}_x}_{\hat{H}_{\textup{QRM}}}+\underbrace{\sum_{i}\Omega_{d}^{(i)}\hat{\sigma}_x\cos{(\omega_{d}^{(i)}t)}}_{\hat{H}_{\textup{drive}}}
    \underbrace{-\omega_q(\beta^{*}\hat{\sigma}_{+}+\beta\hat{\sigma}_{-})-i(\dot{\beta}\hat{\sigma}_{-}-\dot{\beta}^{*}\hat{\sigma}_{+})}_{\hat{H}_{1}}\\& 
    +\underbrace{\sum_{i}\Omega_{d}^{(i)}\cos{(\omega_{d}^{(i)}t)} (\beta^{*}+\beta)\hat{\sigma}_z-\omega_{q}|\beta|^{2}\hat{\sigma}_z -i\frac{1}{2}(\beta^{*}\dot{\beta}-\beta\dot{\beta}^{*})\hat{\sigma}_{z}}_{\hat{H}_{z}}\\ &\underbrace{-g(|\beta|^{2}+\beta^{*})\hat{a}^\dagger\hat{\sigma}_{+}-g(|\beta|^{2}+\beta)\hat{a}\hat{\sigma}_{-}-g(|\beta|^{2}+\beta^{*})\hat{a}\hat{\sigma}_{+}-g(|\beta|^{2}+\beta)\hat{a}^\dagger\hat{\sigma}_{-}}_{\hat{H}_{sb}}\\&\underbrace{+g(\beta^{*}+\beta)(\hat{a}+\hat{a}^\dagger)\hat{\sigma}_z}_{\hat{H}_2}\underbrace{-\sum_{i}\Omega_{d}^{(i)}\cos{(\omega_{d}^{(i)}t)}\beta^{*}(\beta^{*}+\beta)\hat{\sigma}_{+}-\sum_{i}\Omega_{d}^{(i)}\cos{(\omega_{d}^{(i)}t)}\beta(\beta^{*}+\beta)\hat{\sigma}_{-}}_{\hat{H}_3}.
\end{split}
\end{equation}
\end{widetext}
The main purpose of the transformation $\hat{U}$ is to eliminate the $\hat{H}_{\textup{drive}}$, the time-dependent off-diagonal element in $\hat{H}$.
For this, we need to chose proper $\beta$ such that $\hat{H}_{\textup{drive}}+\hat{H}_1=0$ satisfies. 
Even doing so, the Hamiltonian is not fully diagonalized. However, the magnitude of the residual off-diagonal components is smaller than $\hat{H}_{\textup{drive}}$ by a factor of $\beta^2$ or smaller. If $\beta \ll 1$ and $\omega_d$ satisfies the matching conditions for the first order sideband interactions, then the effects from the residual off-diagonal terms other than $\hat{H}_{sb}$ become negligible.
$\hat{H}_z$ accounts for the qubit's frequency shifts and modulations. $\hat{H}_{sb}$ is related with the sideband interactions. $\hat{H}_2$ is derived longitudinal coupling between the qubit and cavity. $\hat{H}_3$ is derived transverse drive. Both $\hat{H}_2$ and  $\hat{H}_3$ are irrelevant to the sideband interaction rates. We neglect the third and higher order terms of $\beta$ in the derivation. We also do not take the dissipative process into consideration in the derivation.

For time-periodical transverse drive, $\beta$ typically takes a form of $\Sigma_{i} \xi_{i} e^{i\omega_d^{(i)}t} + \Sigma_{i}\zeta_{i} e^{-i\omega_d^{(i)}t}$, and here $\xi_{i}$ and $\zeta_{i}$ are time-independent values that we need to find to perturbatively diagonalize the Hamiltonian. Consequently, we can always find the terms 
corresponding to the qubit's frequency modulation in $\hat{H}_z$.
It is intriguing to point out that we obtain the longitudinal drive effect although we start with only the transverse drive fields. We call this derived longitudinal drive in this paper.
The effect of these derived frequency modulation in sideband interaction rates was neglected in many previous works \cite{Blais-PPA-2006,Deppe-NPhys-2008,Zeytinoglu-PPA-2016,Chen-PPA-2016,Krinner-PRApplied-2020,Ann-PRR-2021}. In this study, however, we will prove that these effects significantly contribute to the sideband interaction rates.

\subsection{Monochromatic drive}
\label{2-2}
In this case, we have a drive Hamiltonian $\hat{H}_{\textup{drive}}=2\epsilon_{d}\cos(\omega_d t)\hat{\sigma}_{x}$. For this, the proper $\beta$ is given by,
\begin{equation}
\label{eq:beta}
    \beta = \frac{\epsilon_d}{\Delta}e^{i\omega_d t} + \frac{\epsilon_d}{\Sigma}e^{-i\omega_d t}.
\end{equation}
Here, $\Delta$ and $\Sigma$ are $\omega_q-\omega_d$ and $\omega_q+\omega_d$ respectively.
With this $\beta$, $\hat{H}_{\textup{drive}}+\hat{H}_1=0$ satisfies. For $\hat{H}_z$, we obtain,
\begin{equation}
\label{eq:mono1}
    \hat{H}_z = \hat{\sigma}_z \times \left [  (\frac{\epsilon_d^2}{\Delta}+ \frac{\epsilon_d^2}{\Sigma})(1+2\cos2\omega_d t) - \frac{2\omega_q \epsilon_d^2}{\Delta\Sigma}\cos2\omega_d t \right ],
\end{equation}
which explains the qubit frequency shifts $\delta\omega_q$ and modulation with an amplitude $\Omega_m = 2\epsilon_m$, as given below. 
\begin{equation}
\begin{split}
\label{eq:mono1}
    &\delta\omega_q \approx 2\frac{\epsilon_d^2}{\Delta}+ 2\frac{\epsilon_d^2}{\Sigma},\\&
    \epsilon_m \approx 2\frac{\epsilon_d^2}{\Delta}+ 2\frac{\epsilon_d^2}{\Sigma}-\frac{2\omega_q \epsilon_d^2}{\Delta\Sigma}.
\end{split}
\end{equation}
The matching frequency can be found by considering $\delta\omega_q$ and the dispersive shift $\pm\chi$ in the qubit and cavity induced by the bare coupling $g$. We can approximate $\chi$ by $g^2/\Delta_{qc}+g^2/\Sigma_{qc}$, where $\Delta_{qc}=\omega_q - \omega_c$ and $\Sigma_{qc}=\omega_q + \omega_c$.
Then, the matching conditions for blue and red sideband interactions are given by,
\begin{equation}
\begin{split}
\label{eq:mono1}
    &2\omega_d = \omega_q+\delta\omega_q+\omega_c+ 2\chi ~~\textup{(blue sideband)},\\
    &2\omega_d = |\omega_q+\delta\omega_q-\omega_c+ 2\chi|~~\textup{(red sideband)}.\\
\end{split}
\end{equation}
Eq.~\ref{eq:mono1} is close-form expression of $\omega_d$ because it exists in both the left-hand and right-hand sides. 

When $\omega_d$ satisfies each blue and red sideband condition, then 
the $\hat{H}_{sb}$ at the qubit and cavity rotating frame is reduced to,
\begin{equation}
\label{eq:mono3}
\begin{split}
    &\hat{H}_{sb} = \\& \overbrace{-g(\frac{\epsilon_d^2}{\Delta^2}+\frac{2\epsilon_d^2}{\Delta\Sigma})}\hat{a}^\dagger\hat{\sigma}_{+} + \textup{h.c.} ~~{\textup{(blue sideband)}}.
    \\& -g(\frac{\epsilon_d^2}{\Delta^2}+\frac{2\epsilon_d^2}{\Delta\Sigma})\hat{a}^\dagger\hat{\sigma}_{-} + \textup{h.c.} ~~{\textup{(red sideband}, \omega_q > \omega_c)}.
    \\& \underbrace{-g(\frac{\epsilon_d^2}{\Sigma^2}+\frac{2\epsilon_d^2}{\Delta\Sigma})}_{\Omega_{sb}^{(0)}/2}\hat{a}^\dagger\hat{\sigma}_{-} + \textup{h.c.} ~~{\textup{(red sideband}, \omega_q < \omega_c)}.
\end{split}
\end{equation}
We define $\Omega_{sb}^{(0)}$ by the interaction rates corresponding to the coefficients in front of the operators in Eq.~\ref{eq:mono3}. In addition to $\Omega_{sb}^{(0)}$, there is additional contribution to the sideband interaction rates resulting from the qubit's frequency modulation at $2\omega_d$ in Eq.~\ref{eq:mono1}.  
If $\omega_d$ satisfies the two-photon sideband interactions, then $2\omega_d$ also automatically satisfies the condition for the first order sideband interactions for both blue and red sideband interactions. This phenomena is analogous to inducing the first order sideband interaction by modulating the flux through the squid loop of the frequency tunable qubits, which was first demonstrated in \cite{Strand-PRB-2013}.
We define the interaction rates from this contribution as $\Omega_{sb}^{(1)}$, which amounts to $-2gJ_1(2\epsilon_m/\Delta_{qc})$ for the red sideband interactions, and $-2gJ_1(2\epsilon_m/\Sigma_{qc})$ for the blue sideband interactions.
Here, $J_n(x)$ is the first kind of Bessel function of order $n$. 
The detail derivation is given in \cite{Beaudoin-PRA-2012,Strand-PRB-2013,Navarrete-PRL-2014}.
We summarize the derivation in Sec.~\ref{parameteric}. Finally, we can define $\Omega_{sb}=|\Omega_{sb}^{(0)}+\Omega_{sb}^{(1)}|$ as the analytically predicted sideband interaction rates. 


\subsection{Bi-chromatic drive}
\label{2-3}
Now, we consider the drive Hamiltonian given by
$\hat{H}_{\textup{drive}}=2\epsilon_{dq}\cos(\omega_{dq} t) \hat{\sigma}_{x} + 2\epsilon_{dc}\cos(\omega_{dc} t) \hat{\sigma}_{x}$. The subscription $dq$ and $dc$ refer to qubit friendly and cavity friendly drives, as depicted in Fig.~\ref{fig:energylevels}.
In this case, we chose $\beta$, as below,
\begin{equation}
\label{eq:bi1}
    \beta = \frac{\epsilon_{dq}}{\Delta_1}e^{i\omega_{dq} t} + \frac{\epsilon_{dq}}{\Sigma_1}e^{-i\omega_{dq} t} + \frac{\epsilon_{dc}}{\Delta_2}e^{-i\omega_{dc} t} + \frac{\epsilon_{dc}}{\Sigma_2}e^{i\omega_{dc} t}.
\end{equation}
Here, $\Delta_1,\Delta_2 = \omega_q-\omega_{dq,dc}$ and $\Sigma_1,\Sigma_2 = \omega_q+\omega_{dq,dc}$, respectively. $\hat{H}_z$ in this case is given by,

\begin{equation}
\begin{split}
\label{eq:bi1-1}
    \hat{H}_z = &\hat{\sigma}_z \times \left[ \right. \frac{\epsilon_{dq}^2}{\Delta_1}+ \frac{\epsilon_{dq}^2}{\Sigma_1} + \frac{\epsilon_{dc}^2}{\Delta_2}+ \frac{\epsilon_{dc}^2}{\Sigma_2} 
    \\
    +& \left(\frac{2\epsilon_{dq}^2}{\Delta_1} 
    + \frac{2\epsilon_{dq}^2}{\Sigma_1}
    - \frac{2\omega_{dq} \epsilon_{dq}^2}{\Delta_1\Sigma_1}\right)\cos2\omega_{dq} t
    \\
    +& \left(\frac{2\epsilon_{dc}^2}{\Delta_2}
    + \frac{2\epsilon_{dc}^2}{\Sigma_2}
    - \frac{2\omega_{dc} \epsilon_{dc}^2}{\Delta_2\Sigma_2}\right)\cos2\omega_{dc} t
    \\
    + &{\epsilon_{dq}}{\epsilon_{dc}}\left ( \frac{1}{\Delta_1}+\frac{1}{\Delta_2}+\frac{1}{\Sigma_1}+\frac{1}{\Sigma_2} \right )\cos{(\omega_{dq}-\omega_{dc})t} \\ +& {\epsilon_{dq}}{\epsilon_{dc}}\left ( \frac{1}{\Delta_1}+\frac{1}{\Delta_2}+\frac{1}{\Sigma_1}+\frac{1}{\Sigma_2} \right )\cos{(\omega_{dq}+\omega_{dc})t}
    \left.  \right ].
\end{split}
\end{equation}

The drive induces the frequency shifts $\delta\omega_q$, as given in Eq.~\ref{eq:bi2}. It also modulates the qubit frequency with angular speeds of $\omega_{dq}-\omega_{dc}$ and $\omega_{dq}+\omega_{dc}$, which are the same with the sideband matching frequencies for the bi-chromatic drive case. The amplitude of the modulations at these frequencies is also given in Eq.~\ref{eq:bi2} as $2\epsilon_m$.
\begin{equation}
\begin{split}
\label{eq:bi2}
    &\delta\omega_q \approx 2\frac{\epsilon_{dq}^2}{\Delta_1}+ 2\frac{\epsilon_{dq}^2}{\Sigma_1} + 2\frac{\epsilon_{dc}^2}{\Delta_2}+ 2\frac{\epsilon_{dc}^2}{\Sigma_2},\\&
    \epsilon_m \approx {\epsilon_{dq}}{\epsilon_{dc}}\left[ \frac{1}{\Delta_1}+\frac{1}{\Delta_2}+\frac{1}{\Sigma_1}+\frac{1}{\Sigma_2}\right].
\end{split}
\end{equation}
Then, the matching conditions are given by,
\begin{equation}
\begin{split}
\label{eq:bi3}
    &\omega_{dq}+\omega_{dc} = \omega_q+\delta\omega_q+\omega_c+ 2\chi ~~\textup{(blue sideband)},\\
    &|\omega_{dq}-\omega_{dc}| = |\omega_q+\delta\omega_q-\omega_c+ 2\chi|~~\textup{(red sideband)}.\\
\end{split}
\end{equation}
Eq.~\ref{eq:bi3} is also close-form expression of $\omega_{dq}$ and $\omega_{dc}$.

As in Sec.~\ref{2-2}, we reduce $\hat{H}_{sb}$ as below when the above frequency matching condition satisfies.
\begin{equation}
\label{eq:bi4}
\begin{split}
    &\hat{H}_{sb} = \\& \overbrace{-g(\frac{2\epsilon_{dq}\epsilon_{dc}}{\Delta_1\Delta_2}+\frac{\epsilon_{dq}\epsilon_{dc}}{\Delta_1\Sigma_2}+\frac{\epsilon_{dq}\epsilon_{dc}}{\Delta_2\Sigma_1})}\hat{a}^\dagger\hat{\sigma}_{+} + \textup{h.c.} \\&~~~~~~~~~~~~~~~~~~~~~~~~~~~~~~~~~~~~~~~~~~~~~~{\textup{(blue sideband)}}.
    \\& -g(\frac{\epsilon_{dq}\epsilon_{dc}}{\Delta_1\Delta_2}+\frac{\epsilon_{dq}\epsilon_{dc}}{\Delta_2\Sigma_1}+\frac{\epsilon_{dq}\epsilon_{dc}}{\Sigma_1\Sigma_2})\hat{a}^\dagger\hat{\sigma}_{-} + \textup{h.c.} \\&~~~~~~~~~~~~~~~~~~~~~~~~~~~~~~~~~~~~~~~~~~~~~~{\textup{(red sideband}, \omega_q > \omega_c)}.
    \\& \underbrace{-g(\frac{\epsilon_{dq}\epsilon_{dc}}{\Delta_1\Delta_2}+\frac{\epsilon_{dq}\epsilon_{dc}}{\Delta_1\Sigma_2}+\frac{\epsilon_{dq}\epsilon_{dc}}{\Sigma_1\Sigma_2})}_{\Omega_{sb}^{(0)}/2}\hat{a}^\dagger\hat{\sigma}_{-} + \textup{h.c.} \\&~~~~~~~~~~~~~~~~~~~~~~~~~~~~~~~~~~~~~~~~~~~~~~{\textup{(red sideband}, \omega_q < \omega_c)}.
\end{split}
\end{equation}
From the above Eq.~\ref{eq:bi4}, we can obtain $\Omega_{sb}^{(0)}$. 
We should also consider the effect from $\epsilon_m$ as in Sec.~\ref{2-2}. $\Omega_{sb}^{(1)}$ in this case takes the same expression as in the monochromatic drive case,  $\Omega_{sb}^{(1)}=-2gJ_1(2\epsilon_m/\Delta_{qc})$ or $-2gJ_1(2\epsilon_m/\Sigma_{qc})$ for the red and blue sideband interactions respectively. The analytically predicted sideband interaction rate is then given by $|\Omega_{sb}^{(0)}+\Omega_{sb}^{(1)}|$.

\subsection{First order sideband interaction induced by longitudinal drives}

\label{parameteric}
The goal in this section is to derive the first order sideband interaction rates induced by derived longitudinal drives. The approach we use here is almost identical with that used in Ref.~\cite{Beaudoin-PRA-2012,Strand-PRB-2013,Navarrete-PRL-2014}.
In Eq.~\ref{eq:qrm_mod_1}, we present a Hamiltonian $\hat{H}_m$ reduced from Eq.~\ref{eq5}. We only capture the terms directly relevant to the $\Omega_{sb}^{(1)}$. The qubit is longitudinally driven by a frequency and amplitude $\omega_m$ and $2\epsilon_m$ respectively.
\begin{equation}
\label{eq:qrm_mod_1}
\begin{split}
    \hat{H}_m =
    \frac{{\omega}_q}{2}\hat{\sigma}_z+\epsilon_m\cos(\omega_m t) \hat{\sigma}_z + {\omega_c} \hat{a}^\dagger \hat{a} + g(\hat{a}^\dagger+\hat{a})\hat{\sigma}_x.
\end{split}
\end{equation}
Applying an unitary transformation  $\hat{U}_m=\exp[{i\frac{\epsilon_m}{\omega_m}\sin(\omega_m t)\hat{\sigma}_z}]$ to the above Hamiltonian eliminates the longitudinal drive term ($\epsilon_m\cos(\omega_m t) \hat{\sigma}_z$), while transforming $\hat{\sigma}_x$ operator in the transverse coupling term like below.
\begin{equation}
\label{eq:qrm_mod_2}
\begin{split}
    \hat{U}_m\hat{\sigma}_{\pm}\hat{U}_m^\dagger =  \hat{\sigma}_{\pm} \exp \left [\pm 2i\frac{\epsilon_m}{\omega_m} \sin(\omega_m t) \right].
\end{split}
\end{equation}
We can expand the exponential term using Jacobi–Anger expansion (Eq.~\ref{eq:qrm_mod_3}),
\begin{equation}
\label{eq:qrm_mod_3}
\begin{split}
     \exp \left [i\frac{2\epsilon_m}{\omega_m} \sin(\omega_m t) \right] = \sum_{n=-\infty}^{n=\infty} J_n(\frac{2\epsilon_m}{\omega_m})\exp(in\omega_m t).
\end{split}
\end{equation}
Here, $J_n(x)$ refers to a $n$-th order of the first kind Bessel function.
At the qubit and cavity rotating frame, the Hamiltonian $\hat{H}_m$ can be eventually reduced to,
\begin{equation}
\label{eq:qrm_mod_4}
\begin{split}
    \hat{H}_m \rightarrow g(\alpha(t)\hat{a}\sigma_{+} + \beta(t)\hat{a}\sigma_{-}) + H.c,
\end{split}
\end{equation}
where $\alpha(t)$ and $\beta(t)$ are given by,
\begin{equation}
\label{eq:qrm_mod_5}
\begin{split}
    \alpha(t) = \sum_{n=-\infty}^{n=\infty} J_n(\frac{2\epsilon_m}{\omega_m})e^{-i(\omega_c-\omega_q-n\omega_m)}\\
    \beta(t)  = \sum_{n=-\infty}^{n=\infty} J_n(\frac{2\epsilon_m}{\omega_m})e^{-i(\omega_c+\omega_q+n\omega_m)}
\end{split}
\end{equation}
For $\epsilon_m \ll \omega_m$, we can neglect higher order components ($|n|>1$) in Eq.~\ref{eq:qrm_mod_5} . When $\omega_c$, $\omega_q$, $\omega_m$, and $n$ satisfy specific conditions, we can eliminate the time dependence in the exponents in Eq.~\ref{eq:qrm_mod_5}. Then, the coefficients in front of operators in Eq.~\ref{eq:qrm_mod_4} can be considered a half of sideband interaction rates $\Omega_{sb}^{(1)}/2$ by longitudinal drives.
For example, when $\omega_q-\omega_c = \omega_m$ and $n=-1$ holds, then the longitudinal drive yields the first order red sideband interaction with a magnitude of $\Omega_{sb}^{(1)} = 2gJ_{-1}(\frac{2\epsilon_m}{\omega_m})$. Using the fact $J_{n}(x)=-J_{-n}(x)$, this result is identical to that given in the main text. If $\omega_q<\omega_c$, then we obtain the red sideband interaction when $n=1$, then the $\Omega_{sb}^{(1)}$ is given by $-2gJ_{1}(\frac{2\epsilon_m}{\omega_m})$. For blue the sideband interactions, we need $\omega_q+\omega_c = \omega_m$ with $n=-1$. $\Omega_{sb}^{(1)}$ for this case is also given by 
$-2gJ_{1}(\frac{2\epsilon_m}{\omega_m})$.
One must be careful on the sign of $\Omega_{sb}^{(1)}$ with respect to $\Omega_{sb}^{(0)}$ given in the main text. Otherwise, it results in significant errors in the analytical predictions of $\Omega_{sb}$.

\subsection{Rotating wave approximation}
\label{2-4}
Under the rotating wave approximation (RWA), $\hat{H}_\textup{drive}$ is approximated to,
\begin{equation}
\label{eq:rwa}
    \hat{H}_\textup{drive}^{\textup{(RWA)}} \approx \sum_{i}\frac{\Omega_{d}^{(i)}}{2}(\hat{\sigma}_+e^{-i\omega_d^{(i)} t} + \hat{\sigma}_{-}e^{i\omega_d^{(i)} t}).
\end{equation}
This amounts to taking $\Sigma_{i} \xrightarrow{} \infty$. The RWA model converges to the full model when we have $\Sigma_{i} \gg \Delta_{i}$.
Here we define, $\Delta_{i}=\omega_q-\omega_{d}^{(i)}$ and $\Sigma_{i}=\omega_q+\omega_{d}^{(i)}$.
However, this condition often breakdowns with circuit QED device parameters \cite{Ann-PRR-2021}. 
When the RWA breakdowns, there are significant contributions from the counter-rotating components of $\hat{H}_{\textup{drive}}$ to $\delta\omega_{q}$, $\epsilon_{m}$, and $\Omega_{sb}$.
In Sec.~\ref{Results}, we compare the analytical calculations based on both full and RWA drive models. We confirm that the calculations based on the full drive model show substantially better agreements to the numerical simulation. More detailed discussions will be provided there.

When taking the RWA in this work, we apply the approximation only to the drive Hamiltonian $\hat{H}_d$. Dropping energy non-conservative terms in the interaction part of the QRM is also considered as the RWA. In this case, the QRM is reduced to a Jynes-Cummings (JC) Hamiltonian. However, this yields to a too loose approximation. For example, we cannot capture blue sideband interaction rate when using JC Hamiltonian. Therefore, we always keep the energy non-conservative interaction terms of the QRM in this work.

\subsection{Comparison to transmon-cavity system}
\label{2-5}

Although the QRM deals with two-level systems coupled to a linear cavity, many qubit systems realized experimentally are not exactly two-level systems. One famous example is transmon qubits widely used nowadays, which can be considered weakly anharmonic Duffing oscillators. 
From Ref.~\cite{Ann-PRR-2021}, the system and monochromatic drive Hamiltonian of a dispersively coupled transmon and cavity system in their normal mode basis can be expressed by,

\begin{equation}
\begin{split}
    &\hat{H}_{\textup{transmon}} \approx ~ \, ({\omega}_{t}+\chi_{t}) \hat{a}^\dagger \hat{a}+ {\omega}_{c} \hat{b}^\dagger \hat{b}  \\
    & ~~ - \frac{1}{12} \left[ \chi_t^{1/4} (\hat{a} + \hat{a}^\dagger) + \chi_c^{1/4} (\hat{b} + \hat{b}^\dagger) \right] ^4.\\
    &\hat{H}_{\textup{drive}} = \,\Omega_d (\hat{a}^\dagger + \hat{a})\cos(\Omega_d t).
\label{transmon1}
\end{split}
\end{equation}
Here, $\hat{a}$ and $\hat{b}$ are transmon and cavity mode destruction operators respectively. $\chi_{t,c}$, and $\omega_{t,c}$ refer to the Duffing nonlinearity and resonant frequencies of the transmon and cavity modes respectively. $\chi_{tc}\approx\sqrt{\chi_{t}\chi_{c}}\sim g^2$ is defined as a cross Duffing nonlinearity.
In this case, the Schriffer-Wollf transformation $\hat{U}(t)=e^{\xi(t)\hat{a}^\dagger-\xi(t)^{*}\hat{a}}$ acting on the Hamiltonian simply displaces $\hat{a}$ to $\hat{a}-\xi$ while eliminating the drive term. $\xi(t)$ is given by $\frac{\Omega_{d}}{2\Delta}e^{-i\omega_{d}t}+\frac{\Omega_{d}}{2\Sigma}e^{i\omega_{d}t}$. Then, the total Hamiltonian $\hat{H}_{\textup{transmon}}+\hat{H}_{\textup{drive}}$ is transformed to $\hat{H}'$.
\begin{equation}
\begin{split}
    \hat{H'} \approx ~ \,& ({\omega}_{t}+\chi_{t}) \hat{a}^\dagger \hat{a}+ {\omega}_{c} \hat{b}^\dagger \hat{b}  \\
    & - \frac{1}{12} \left[ \chi_t^{1/4} (\hat{a} + \hat{a}^\dagger - \xi(t) -\xi^{*}(t)) + \chi_c^{1/4} (\hat{b} + \hat{b}^\dagger) \right] ^4.
\label{transmon2}
\end{split}
\end{equation}
The weakly anharmonic nature of the transmon dramatically simplifies the analytical derivation.
The qubit's frequency shifts, modulation, and the sideband interaction rates are captured in the fourth power term in Eq.~\ref{transmon2} like below,

\begin{equation}
\begin{split}
   \delta\omega_{t} = ~ -\frac{1}{2}\Omega_{d}^{2}\chi_{t}\times (\frac{1}{\Delta^2} +\frac{2}{\Delta\Sigma}+\frac{1}{\Sigma^2}),\\
    \epsilon_{m} = ~ -\frac{1}{2}\Omega_{d}^{2}\chi_{t}\times (\frac{1}{\Delta^2} +\frac{2}{\Delta\Sigma}+\frac{1}{\Sigma^2})
   \\
    \Omega_{sb} = ~ -\frac{1}{2}\Omega_{d}^{2}\chi_{t}^{1/2}\chi_{tc}^{1/2}\times (\frac{1}{\Delta^2} +\frac{2}{\Delta\Sigma}+\frac{1}{\Sigma^2}),
\label{transmon3}
\end{split}
\end{equation}
which differ from those in the case of the QRM. $\delta\omega_t$ is proportional to $\Omega_d^2/\Delta^2$ in the RWA regime, showing a different form compared with Eq.~\ref{eq:mono1}. Furthermore, we can confirm the collusion effect between the co-rotating and counter-rotating terms in $\delta\omega_t$ in Eq.~\ref{transmon3} that does not appear in $\delta\omega_q$ in Eq.~\ref{eq:mono1}. Qualitative different feature can be also found in $\Omega_{sb}$. In the transmon case, although we can also have derived longitudinal drives with a frequency of $2\omega_t$, it does not contribute to the sideband interaction rates unlike the driven QRM case.

\section{Benchmarking with numerical simulations}
\label{Results}

\subsection{Overview}
\label{Overall}
To verify the validity of the derived formula, we perform the numerical simulation with several system parameter sets. 
We define the drive and transition frequencies of the qubit and cavity as $f_{d}=\omega_{d}/2\pi$, $f_{q}=\omega_{q}/2\pi$, and $f_{c}=\omega_{c}/2\pi$, respectively.
In the bi-chromatic drive case, we define the qubit and resonator drive frequencies as $f_{dq}=\omega_{dq}/2\pi$ and $f_{dc}=\omega_{dc}/2\pi$, respectively.
For the QRM parameters, we investigate two cases here: $f_{q,c}=6.5, 4.0$ GHz and $f_{q,c}=4.0, 6.5$ GHz.

In the numerical simulation, we solve the time-dependent master equation of the driven QRM, and we get the time evolution of the qubit and cavity. 
For monochromatic drive cases, we sweep the $f_{d}$ until the resonant sideband interaction takes place to find the matching frequencies. The procedure is somewhat complicated for bi-chromatic drive cases. First, we fix $f_{dc}$ by $f_{c}-500$ MHz. We parameterize the $\epsilon_{dq}/2\pi$ and $\epsilon_{dc}/2\pi$ with a real positive parameter $\eta$. Both are given by $\epsilon_{dq}/2\pi=\eta\cdot25$ MHz and $\epsilon_{dc}/2\pi=\eta\cdot317$ MHz, respectively. With these conditions, we sweep the $f_{dq}$ until the resonant sideband interaction takes place. 
More detail of the procedure for the numerical simulation is given in Sec.~\ref{simulation}.

\begin{figure}
    \centering
    \includegraphics[width=1\columnwidth]{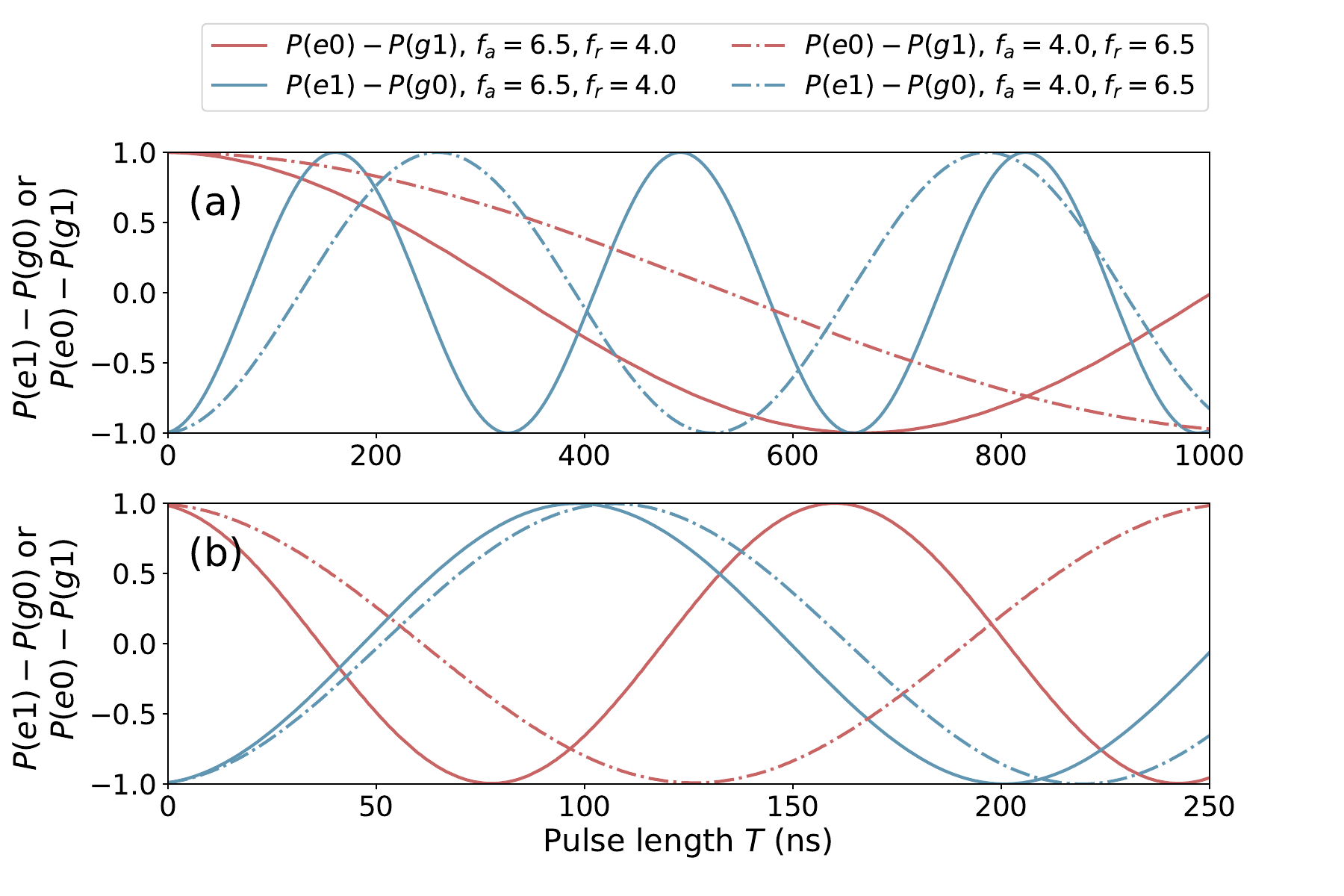}
    \caption{Time-domain numerical simulation results of the first order two-photon sideband interactions in the QRM with various system configurations. Eight different cases are present. The red and blue lines indicate the red and blue sideband interactions, respectively. (a) The sideband interactions by the monochromatic drive fields. (b) The sideband interactions by bi-chromatic drive fields. Please see the text and legend for further detail of the conditions in the numerical simulation.
    }
    \label{fig:timedomain}
\end{figure}

In Fig.~\ref{fig:timedomain}, we present the simulated time domain dynamics when the two-photon sideband interactions take place. $g/2\pi$ in both cases is fixed by 200 MHz. Fig.~\ref{fig:timedomain}(a) shows the results under a monochromatic drive with $\epsilon_d/2\pi=300$ MHz. Fig.~\ref{fig:timedomain}(b) shows the results under a bi-chromatic drive when $\epsilon_{dq}/2\pi=25$ MHz and $\epsilon_{dc}/2\pi=317$ MHz, respectively ($\eta=1$). The definitions of $\epsilon_d$, $\epsilon_{dq}$, and $\epsilon_{dc}$ are the same as in the previous section. We can also confirm that whether the qubit is red or blue detuned to the cavity results in different sideband interaction rates. This is already predictable from the analytical formula derived in Sec.~\ref{theory}. 
We do not introduce any dissipative process in the numerical simulation. Unless the dissipation rates become comparable to the sideband interaction rates, there is no noticeable change in the sideband oscillation frequency of the time-domain numerical simulations.

In this section, we plot the numerical simulation results with analytical predictions with four different models. The models with a full drive Hamiltonian are labeled by `Full' in the legend, whereas models with the RWA in the drive Hamiltonian are labeled by `RWA'. We also separately plot the results with and without considering the effect of the derived longitudinal drives (labeled by $|\Omega_{sb}^{(0)}+\Omega_{sb}^{(1)}|$ and $|\Omega_{sb}^{(0)}|$ respectively).
The model used in Ref.~\cite{Blais-PPA-2006} corresponds to the results labeled by `RWA, $|\Omega_{sb}^{(0)}|$' in this paper.

\subsection{Method for numerical simulation}
\label{simulation}
The dynamics of the system can be described by the equation,
\begin{equation}
\label{eq:master}
\begin{split}
    d\hat{\rho}_{sys}/dt = -i[\hat{H}_{\textup{QRM}}+\hat{H}_{\textup{drive}},\hat{\rho}_{sys}].
\end{split}
\end{equation}
Here, $\hat{\rho}_{sys}$ is a density matrix of the qubit and cavity. We do not take the dissipation into consideration. 
In the numerical study in this paper, we rigorously benchmark the real experiments. We set the rising and falling in the sideband drive strength as in the real experiments. Specifically, $\epsilon_{d}(t)$ is defined as a pulse with 10-ns of Gaussian rising and falling time.
We can then scan the pulse length and plot the quantum states of the system at the end of each pulses. We do not include the rising and falling times in the definition of the pulse length. 

\begin{figure}
    \centering
    \includegraphics[width=1\columnwidth]{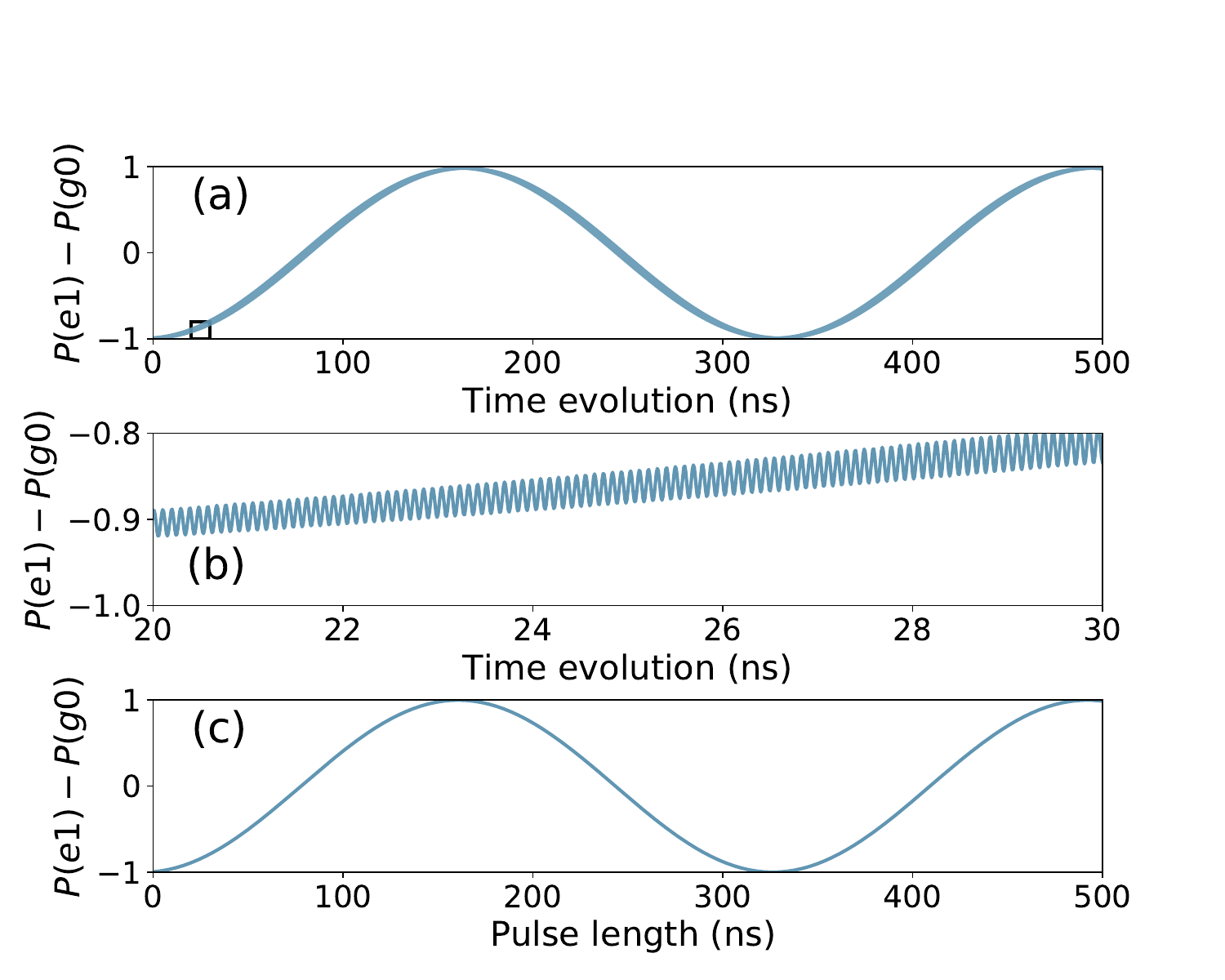}
    \caption{Time-domain numerical simulation. 
    (a) A direct solution of the master equation when the drive field satisfies the blue sideband interaction. Please see the text for a description of the conditions in the simulation. 
    We consider 10 ns Gaussian rising and falling time in the drive amplitude.
    (b) Zoom in on the black rectangular box in (a). We identify that the fast micro-oscillation and the frequency of this oscillation is the same as the drive frequency.
    (c) We plot the $P(e1)-P(g0)$ at the end of the pulse with respect to the pulse length without rising and falling times. A clear sinusoidal oscillation is obtained. 
    }
    \label{fig:convergent}
\end{figure}

Fig.~\ref{fig:convergent} provides a step-by-step description of our numerical simulation method. The simulation parameters used in Fig.~\ref{fig:convergent} are $\omega_q,\omega_c,\omega_d, \epsilon_d, g$ = $2\pi\times (6.5,4.0,0.1,0.2,5.278)$ GHz. The monochromatic drive frequency $\omega_d$ satisfies the matching condition for the blue sideband interaction.
Fig.~\ref{fig:convergent}a shows the dynamics of the system under the sideband drive pulse with a length of 480 ns.
Fig.~\ref{fig:convergent}(b) magnifies the area enclosed by the square in Fig.~\ref{fig:convergent}(a). One can identify the fast but small oscillation in the quantum state of the system. This oscillation originates from the Hamiltonian's time dependence. We can remove the time dependence by moving to the rotating frame at $\omega_d$, and removing all the fast rotating components. This is what amounts to the rotating wave approximation (RWA). However, the RWA is only available when the $\epsilon_d$ and $|\omega_q-\omega_d|$ are small enough. These conditions are clearly not satisfied for the two-photon sideband interaction with circuit QED parameters.
We repeat the simulations by varying the pulse lengths, and we plot the states at the end of the pulses (when the pulse falling finishes). The result is given in Fig.~\ref{fig:convergent}(c). We obtain a clear sinusoidal curve without the fast oscillation.

The procedure described above is analogous with the real experiment. This explains why one still can see clear sinusoidal dynamics in the experiment, even with a very strong drive strength. We calculate $P(e1)-P(g0)$ for the blue sideband interactions, and $P(e0)-P(g1)$ for the red sideband interactions in this paper. Here, $P$ refers to the probability to find the system in the states enclosed in the brackets. Once we obtain a sinusoidal oscillation from the simulation, we then determine the sideband interaction rate $\Omega_{sb}$ from the period of the oscillation.

\begin{figure}
    \centering
    \includegraphics[width=1\columnwidth]{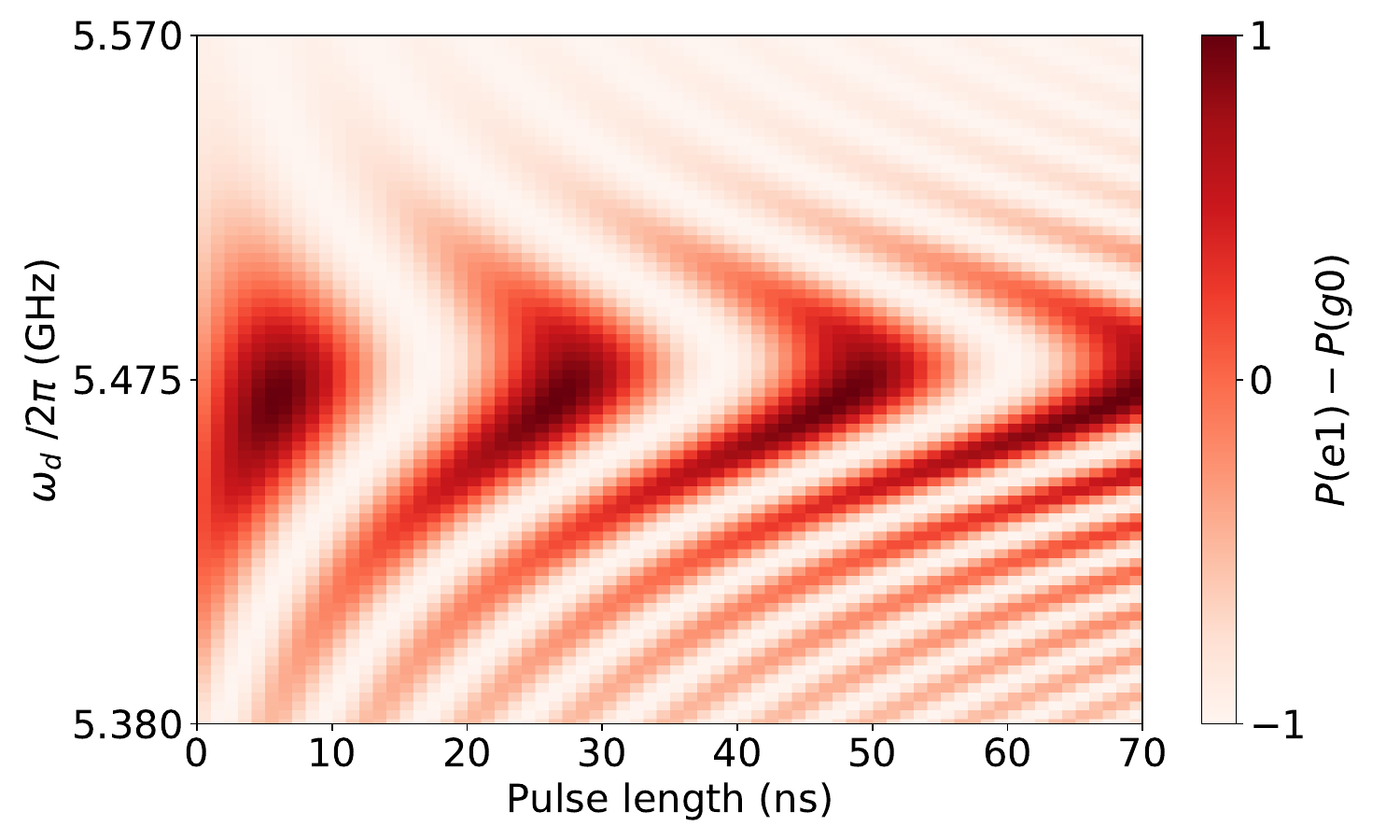}
    \caption{Finding a matching frequency.
    This plot shows the dynamics of the two-state system (qubit) when we sweep the monochromatic drive frequency around the matching frequency that satisfies the resonant blue sideband interaction. Please see the text for a description of the conditions that we used in the simulation. 
    }
    \label{fig:chevron}
\end{figure}
Fig.~\ref{fig:chevron} describes how we found the matching frequencies for sideband interactions. All of the simulation parameters are the same in Fig.~\ref{fig:convergent} except that $\epsilon_d/2\pi$ is 500 MHz.
We sweep $\omega_d$ around the predicted matching frequency for blue sideband interaction.
In this case, the matching frequency is found at $\omega_d/2\pi=5.474$ GHz. The asymmetric shape is attributed to the fact that the frequency shift of qubit changes while sweeping the drive frequency. 

\subsection{Monochromatic drives}
\label{monochromatic}
\begin{figure}
    \centering
    \includegraphics[width=1\columnwidth]{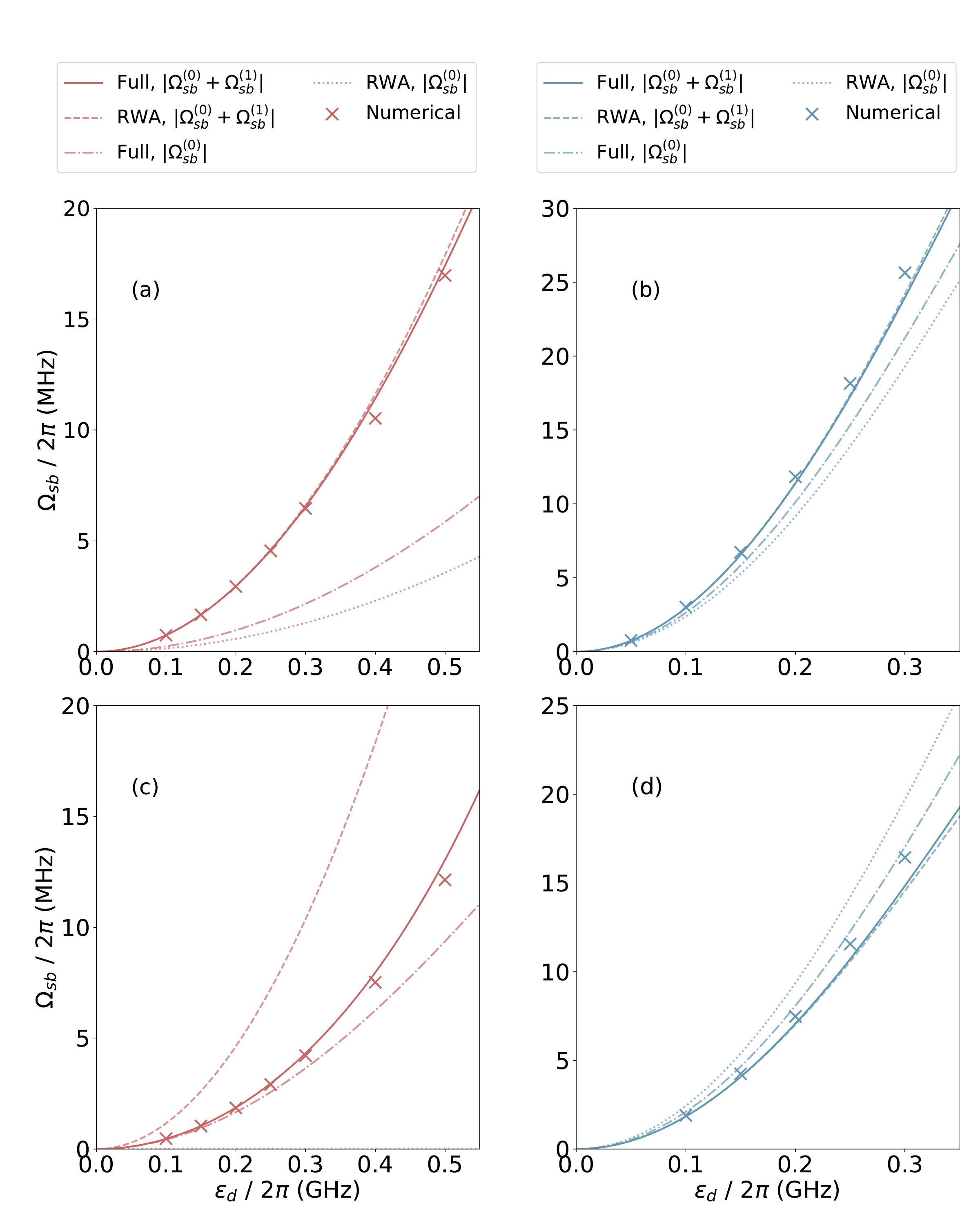}
    \caption{Red and blue sideband interaction rates ($\Omega_{sb}$) induced by monochromatic drive fields. The lines indicate the analytically calculated sideband interaction rates. See the legend for the detail information. The cross marks indicate the numerically simulated results based on the $\hat{H}+\hat{H}_{drive}$. (a,b) $f_q$ = 6.5 GHz and $f_c$ = 4.0 GHz. (c,d) $f_q$ = 4.0 GHz and $f_c$ = 6.5 GHz.  In the case of (c), the fine dashed line (RWA, $\Omega_{sb}^{(0)}$ in the legend) lies on x-axis, and thus hardly visible in the figure.
    }
    \label{fig:single-sbrates}
\end{figure}

In this subsection, we deal with only the monochromatic drive cases.
The parameters that we use for the simulations are $f_{q,c}=6.5, 4.0$ GHz in Fig.~\ref{fig:single-sbrates}(a,b)  and $f_{q,c}=4.0, 6.5$ GHz in Fig.~\ref{fig:single-sbrates}(c,d). 
$g/2\pi$ in both cases is fixed by 200 MHz.
The lines in Fig.~\ref{fig:single-sbrates} show the the analytically calculated sideband interaction rates. We first obtained the matching frequencies based on Eq.~\ref{eq:mono1}, and we use these values when calculating the sideband interaction rates.
When analytically calculating $\Omega_{sb}$ here and in the following of this paper, we replace $\omega_q$ in the formula with $\omega_q+\delta\omega_q$ for higher accuracy.

In Fig.~\ref{fig:single-sbrates}, the sideband interaction rates calculated by the full drive model with the derived longitudinal drive (solid lines) excellently agree with all the numerical simulation results, whereas the other model fails to explain all parameter cases. Noticeably, the derived longitudinal drive ($\Omega_{sb}^{(1)}$) significantly accounts for the total sideband interaction rates. All these trends can be also found in the bi-chromatic drive cases in Sec.~\ref{bichromatic}.

As $\epsilon_d$ becomes larger, the accuracy of the analytical model decreases. This happens because the basic assumption for perturbative approach ($\epsilon_{d}/|\omega_{q}-\omega_{c}|\ll1$) in derivation of the analytical model becomes weakened. We can understand the large discrepancy in blue sideband cases in the same manner. The blue sideband interaction requires the matching frequency $f_d$ much closer to the $f_q$ than the red sideband interaction does.
In Fig.~\ref{fig:single-sbrates}(d), the numerical results with large drive strengths are more consistent with another analytical model (double-dashed line) rather than full model (solid line). This is an interesting coincidence to point out. The sideband interactions contributed by derived longitudinal drive is significant in the red sideband cases but not in the blue sideband cases.

\subsection{Bi-chromatic drives}
\label{bichromatic}
\begin{figure}
    \centering
    \includegraphics[width=1\columnwidth]{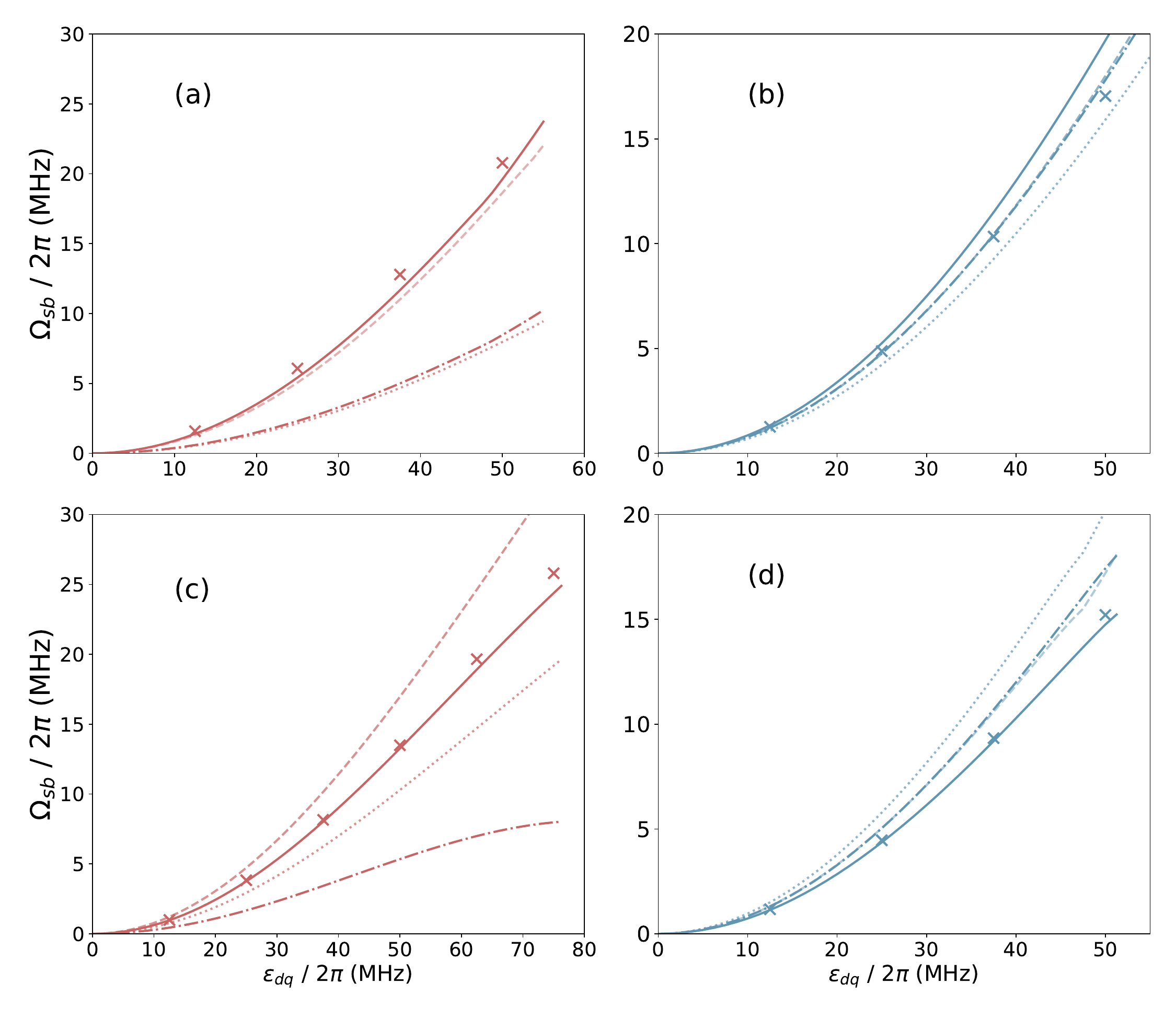}
    \caption{Red and blue sideband interaction rates $\Omega_{sb}$ induced by bi-chromatic drive fields.
    The lines indicate the analytically calculated sideband interaction rates. The cross marks indicate the numerically simulated results based on the $\hat{H}+\hat{H}_{drive}$.
    See the legend in Fig.~\ref{fig:single-sbrates} for more detailed information. $\omega_{dc}$ is fixed by $f_{c}-500$ MHz. $\epsilon_{dq}$ and $\epsilon_{dc}$ are parameterized as described in Sec.~\ref{simulation} (a,b) $f_q$ = 6.5 GHz, $f_c$ = 4.0 GHz. (c,d) $f_q$ = 4.0 GHz, $f_c$ = 6.5 GHz.
    }
    \label{fig:double-sbrates}
\end{figure}

We investigate the bi-chromatic drive cases in this subsection. 
$f_{q,c}$ and $g$ used in the simulations are the same as in Sec.~\ref{monochromatic}.
We analytically find the proper $f_{dq}$ based on the Eq.~\ref{eq:bi2}, fixing the $f_{dc}$ to $f_{c}-500$ MHz. $\epsilon_{dq}$ and $\epsilon_{dc}$ are parameterized as described in Sec.~\ref{simulation}. We also analytically calculate the sideband interactions based on the results in Sec~\ref{Overall}.  
Fig.~\ref{fig:double-sbrates} compares the sideband interaction rates calculated by the numerical simulation (cross) and analytical calculation (line).
by the full drive model with the derived longitudinal drive (solid lines) explains the simulation results better than other models, except for the one case in Fig.~\ref{fig:double-sbrates}(b). 
In the red sideband cases, we can clearly see the significant effect of the derived longitudinal drive in the sideband interaction rates ($\Omega_{sb}^{(1)}$). This also results in approximately 15\% correction to the total sideband interaction rates in the blue sideband cases.
The effect of the RWA is very conspicuous in Fig.~\ref{fig:double-sbrates}(c) but not in other cases. In particular, we can hardly identify the effect of the RWA in Fig.~\ref{fig:double-sbrates}(a).


\subsection{From strong to ultrastrong coupling regime}
In the previous subsections, we have fixed $g/2\pi$ by 200 MHz. In this subsection, we perform the simulation with different $g$ while fixing the drive strengths and the other system parameters. We use $f_q$ = 4.0 GHz and $f_c$ = 6.5 GHz in the simulation. We scan $g$ from 100 MHz (strong coupling regime) to 500 MHz (ultrastrong coupling regime). 

\begin{figure}
    \centering
    \includegraphics[width=1\columnwidth]{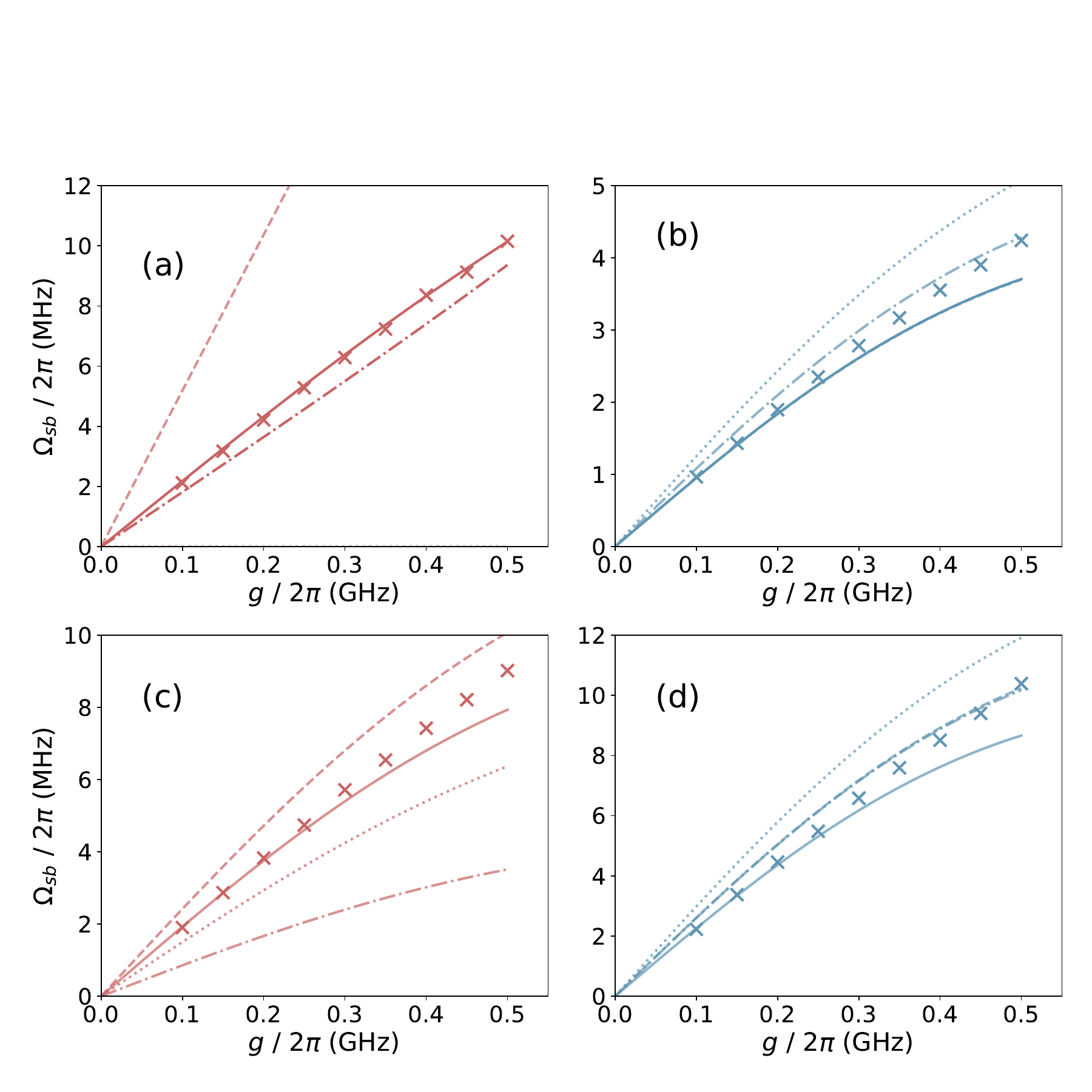}
    \caption{Red and blue sideband interaction rates $\Omega_{sb}$ with different qubit—cavity coupling strength $g$. We fix $f_q$ = 4.0 GHz and $f_c$ = 6.5 GHz in the calculation.
    The lines indicate the analytically calculated sideband interaction rates. The cross marks indicate the numerically simulated results based on the $\hat{H}+\hat{H}_\textup{drive}$.
    See the legend in Fig.~\ref{fig:single-sbrates} for the detail information.
    (a,b) Monochromatic drive cases. Drive strength is fixed by $\epsilon_d/2\pi$ = 100 MHz. 
    (c,d) Monochromatic drive cases. Drive strength is fixed by $f_{dc}$ is fixed by $f_{c}-500$ MHz. In all cases, we set $\epsilon_{dq}/2\pi$ = 25 MHz and $\epsilon_{dc}/2\pi$ = 317 MHz. In the case of (a), the fine dashed line (RWA, $\Omega_{sb}^{(0)}$ in the legend) lies on x-axis, and thus hardly visible in the figure.}
    \label{fig:gscan}
\end{figure}

In Fig.~\ref{fig:gscan}, 
we plot the red and blue sideband interaction rates with different qubit–cavity coupling strength $g$. Fig.~\ref{fig:gscan}(a,b) describe mono-chromatic drive cases and Fig.~\ref{fig:gscan}(c,d) describe bi-chromatic drive cases.
Drive strengths $\epsilon_{d}$ are fixed by 100 MHz (red) and 300 (blue) MHz, respectively.  
Similar to the previous results in Fig.~\ref{fig:single-sbrates} and Fig.~\ref{fig:double-sbrates}, 
the full drive model with the derived longitudinal drive (solid lines) explains the numerical simulation results better than the other model does when the drive strengths are small enough. As $g$ becomes larger, the discrepancy between the numerical and analytical values also becomes larger.
Eventually, the numerical results fall into the other analytical models in (b) and (d). We can also confirm that a significant portion of the $\Omega_{sb}$ is attributed to $\Omega_{sb}^{(1)}$.

\section{Conclusion}
\label{conclusion}
In this paper, we have analytically and numerically studied the first order sideband interactions that are induced by two-photon drives in a quantum Rabi Hamiltonian. We confirm that the sideband interaction rates can be accurately predicted based on the analytical formula when the parameters are in the perturbative regime ($\epsilon_d/|\omega_q-\omega_d|$). We also confirm that the RWA significantly misleads the prediction of the sideband interaction rates for some system parameters. 
We also find that the transverse drive field can induce the derived longitudinal drive Hamiltonian. In addition, we can confirm its significant contributions to total sideband interaction rates.  
As the drive parameters deviate from the perturbative regime, we observe disagreement between numerical and analytical calculation, and consequently the other models coincidentally provide more accurate predictions.  
Our study significantly improves the accuracy of the analytical formula from the previous work. 

\begin{acknowledgements}
We acknowledge David Vitali for our helpful discussions.
Byoung-moo Ann also acknowledges support from the European Union’s Horizon 2020 research and innovation program under the Marie Sklodowska-Curie grant agreement No. 722923 (OMT).
This project also has received funding from the European Union’s Horizon 2020 research and innovation program under grant agreement No. 828826 - Quromorphic. The data that support the findings of this study are available in \cite{data}.

Author contribution : 
B.A conceived the study and made the theoretical description. W.K developed the numerical simulation software under the supervision of B.A. B.A analyzed data and wrote the manuscript. G.A.S supervised the study.
\end{acknowledgements}

\appendix

\section{Extended data: Estimation of the matching frequencies}
\label{appendix-1}
\begin{figure}
    \centering
    \includegraphics[width=1\columnwidth]{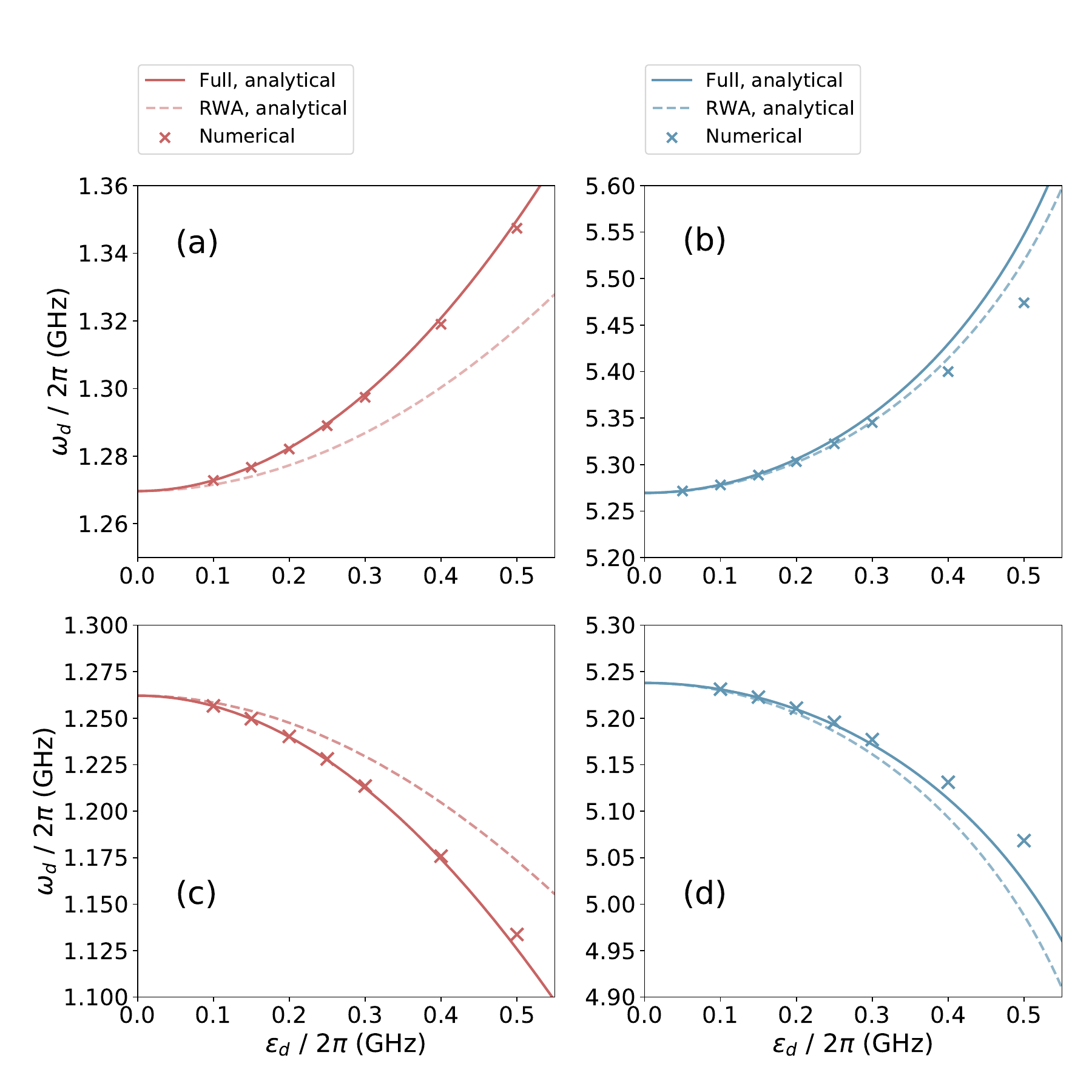}
    \caption{Matching drive frequencies ($\omega_{d}$) for two-photon red and blue sideband interactions induced by monochromatic drive fields. The single and double dashed lines indicate the analytically calculated matching frequencies the red and blue sideband interactions respectively. These are based on the full (single-dashed) and RWA model (double-dashed). The cross marks indicate the numerically simulated results based on the full model. (a,b) $f_q$ = 6.5 GHz and $f_c$ = 4.0 GHz. (c,d) $f_q$ = 4.0 GHz and $f_c$ = 6.5 GHz.
    }
    \label{fig:single-shifts}
\end{figure}
\begin{figure}
    \centering
    \includegraphics[width=1\columnwidth]{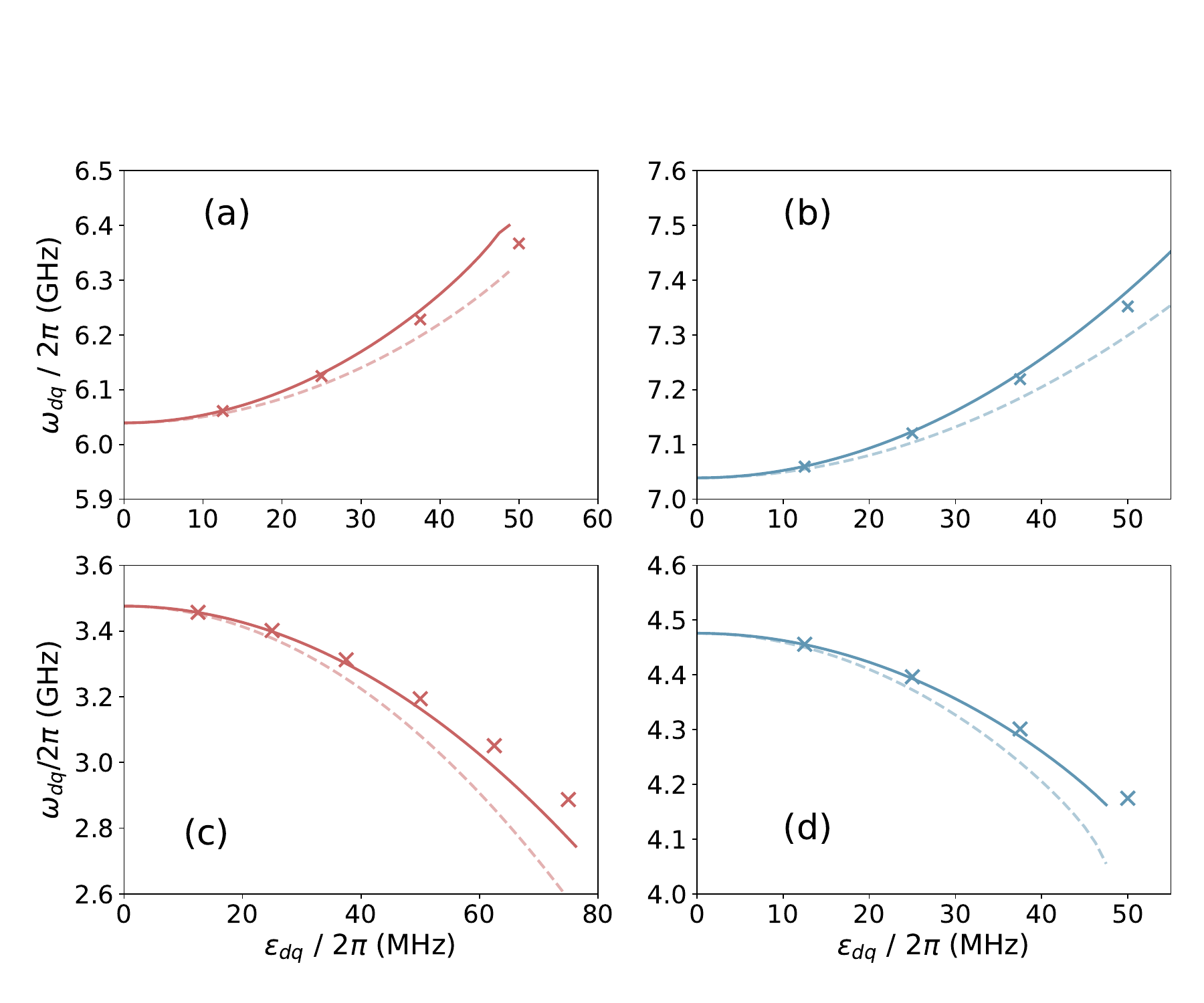}
    \caption{Matching drive frequencies ($\omega_{dq}$) for two-photon red and blue sideband interaction induced by bi-chromatic drive fields when $f_{dc}$ is fixed by $f_{c}-500$ MHz. 
    $\epsilon_{dq}$ and $\epsilon_{dc}$ are parameterized as described in Sec.~\ref{simulation}. The single and double dashed lines indicate the analytically calculated matching frequencies of the red and blue sideband interactions, respectively. These are based on the full (single-dashed) and RWA model (double-dashed). The cross marks indicate the numerically simulated results based on the full model. (a,b) $f_q$ = 6.5 GHz and $f_c$ = 4.0 GHz. (c,d) $f_q$ = 4.0 GHz and $f_c$ = 6.5 GHz.
    }
    \label{fig:double-shifts}
\end{figure}
\begin{figure}
    \centering
    \includegraphics[width=1\columnwidth]{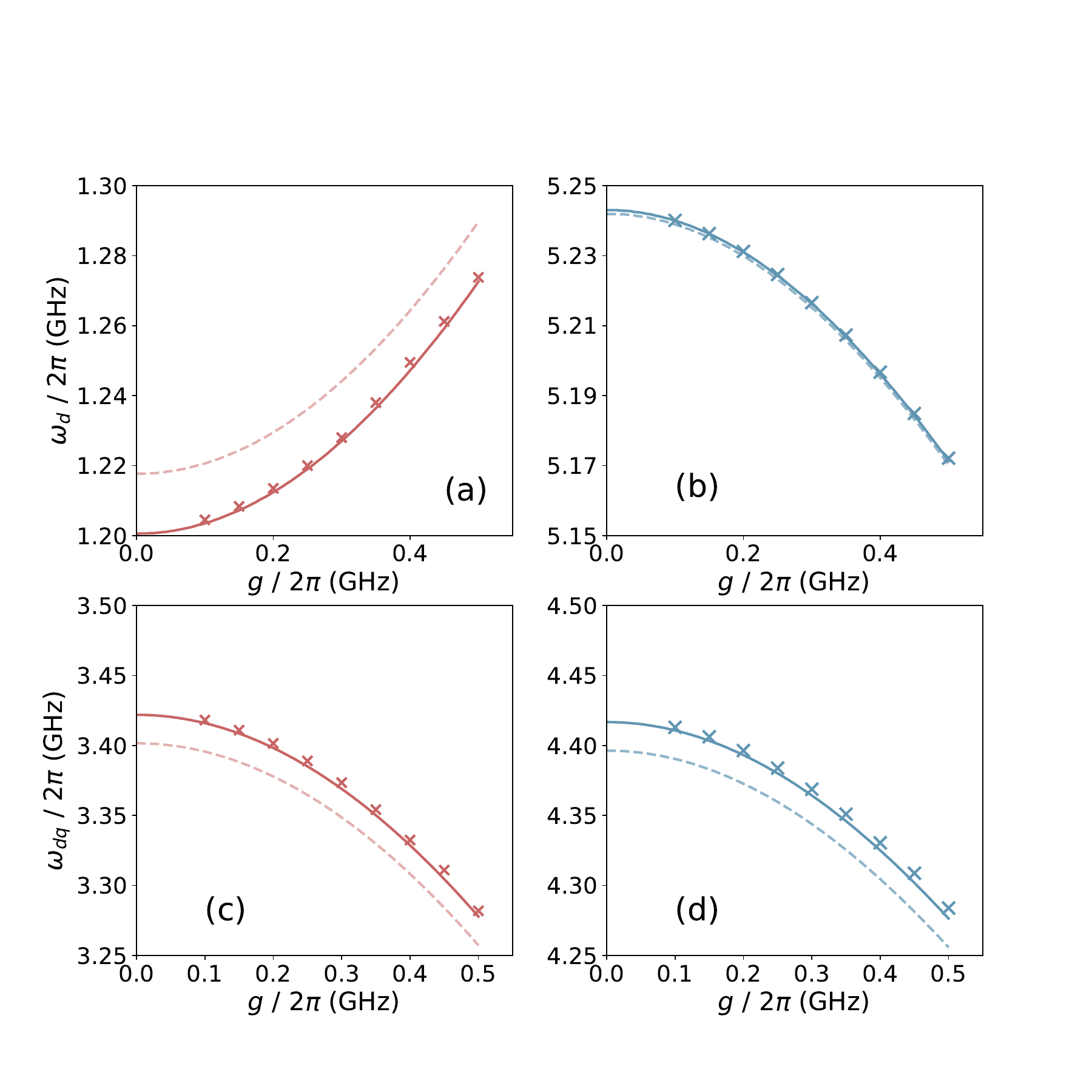}
    \caption{The matching drive frequencies with different qubit and cavity bare coupling $g$. $f_q$ = 4.0 GHz and $f_c$ = 6.5 GHz in the calculation. (a,b) Mono-chromatic drive cases. Drive strength is fixed by $\epsilon_d/2\pi$ = 100 MHz. (c,d) Bi-chromatic drive cases.  $f_{dc}$ is fixed by $f_{c}-500$ MHz. In all cases, we set $\epsilon_{dq}/2\pi$ = 25 MHz and $\epsilon_{dc}/2\pi$ = 317 MHz.
    }\label{fig:gscan-matching}
\end{figure}
In this section, we present the analytically calculated matching frequencies compared to the numerical simulation results. All of the simulation conditions and parameters are the same in Fig.~\ref{fig:single-sbrates}, Fig.~\ref{fig:double-sbrates}, and Fig.~\ref{fig:gscan}.
The lines in Fig.~\ref{fig:single-shifts}, Fig.~\ref{fig:double-shifts}, and Fig.~\ref{fig:gscan-matching}  show the analytically calculated matching frequencies based on the formula that we obtained in Sec.~\ref{2-2}. The matching frequencies obtained from the numerical simulation are denoted by cross marks.
In general, the calculated matching frequencies based on the full drive Hamiltonian are in better agreement with the numerical results. The only the exception is Fig.~\ref{fig:single-sbrates}(b). In this case, the numerical data deviates from the full analytical model due to the breakdown of the perturbative approach, and eventually gets closer to the RWA model coincidentally.

\end{document}